\documentclass[12pt]{article}

\usepackage{moriond}
\usepackage{astro}
\usepackage{latexsym}
\usepackage{epsfig}

\begin{document}

\mathversion{bold}

\heading{Star formation at high redshift as traced by near-infrared
H$\alpha$ emission surveys}

\mathversion{normal}

\makeatletter
\renewcommand{\@makefnmark}{\mbox{\ }}
\makeatother

\footnote{Invited Review to apper in {\it
``Extragalactic Astronomy in the Infrared''}, eds.\ G.\ A.\ Mamon,
Trinh Xu\^an Thu\^an, \& J.\ Tr\^an Thanh V\^an, Editions
Fronti\`eres, Gif-sur-Yvette}

\renewcommand{\author}[3]{
        \vspace{5mm}  
        \begin{center}
                {\normalsize \rm #1}\\    
                {\normalsize \it #2}\\    
                {\normalsize \it #3}\\    
                \vspace{0.65cm}
                \epsfig{file=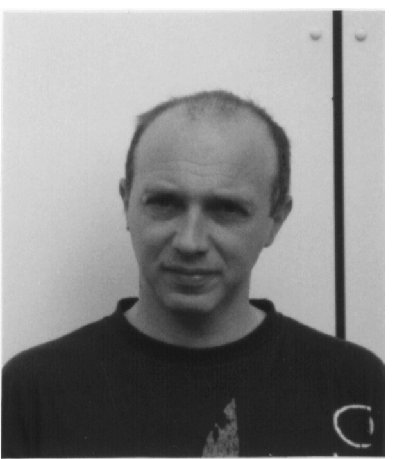,height=3.8cm}
                \vspace{0.35cm}
        \end{center}
}

\author{Paul P. van der Werf} {Leiden Observatory, The Netherlands.}  {\ }

\begin{moriondabstract}
Surveys of redshifted emission lines in the near-infrared
will provide unique information on the cosmic star
formation history. Near-infrared $\Ha$ surveys will probe the cosmic
star formation density and the evolution of the luminosity function of
star forming galaxies out to $z\sim2.5$, including the important
$z=1.5-2.0$ epoch, where according to current knowledge the cosmic
star formation density peaks. $\Ha$
and other hydrogen recombination lines (except $\Lya$) and
[$\ion{O}{ii}$] 
are all useful tracers of star formation 
provided the luminosity function is probed to sufficiently deep
levels. However, shallow surveys or surveys of $\Lya$ or
[$\ion{O}{iii}$] emission are not suitable for measuring or even
constraining the cosmic star formation density.
A first estimate of the star formation
density at $z\approx2.25$ is derived, based on a new and sensitive large-area
ESO near-infrared survey for $\Ha$ emission
at this redshift. Future observational directions, at near-infrared
and longer wavelengths, for reading the cosmic star formation history
are outlined.
\end{moriondabstract}

\pagestyle{plain}

\section{Introduction}\label{sec.Intro}

The very rapid developments in near-infrared (near-IR) array technology of the
last few years have brought about a revolution in the field of
emission line surveys in the near-IR wavebands. Until
recently such surveys did not yield any detections 
\cite{Thompsonetal94,PahreDjorgovski95,Bunkeretal95}, although these
null-results in some cases
began to set interesting limits on the number density of the most
luminous emission line galaxies \cite{Thompsonetal96}. The recent 
availability of near-IR cameras with large-format arrays has now produced
the first detections in surveys for $\Ha$ emission at high
redshift. While the number of detections is still small, this result
for the first time provides a crude estimate of the
density of luminous emission line objects at cosmological distances.
Simultaneously, a change of focus has taken place. Whereas previously
surveys were aimed at finding only the most luminous objects, current
technology allows deeper searches, and for the first time an estimate
of the star formation density (SFD) at $z\approx2.25$ is obtained.
Future surveys will be able to probe the luminosity function to even
greater depth.

This review first discusses our current state of knowledge of the star
formation history of the universe, and the crucial role
which near-IR emission line surveys will play in this
field (\secref{sec.SFD}). Subsequently the use (and limitations) 
of emission lines as cosmic star formation
tracers is discussed
(\secref{sec.lines}). Then,
results of the ESO near-IR $\Ha$ survey are presented, and
implications for the cosmic SFD at $z\approx2.25$ are
derived (\secref{sec.surveys}). 
Finally, prospects for future observational work, using $\Ha$
and other tracers, for tracing the star formation history of the
universe are discussed (\secref{sec.con}).
Throughout this paper the Hubble constant is written as
$H_0=75\,h_{75}\kms\pun{Mpc}{-1}$ and $q_0=0.1$ is assumed.

\section{Cosmic star formation history}\label{sec.SFD}

The star formation properties of the universe at high redshift form one
of the focal points of current observational research in cosmology.
Recently, our current understanding of the cosmic star formation
history has been summarized
\cite{Madauetal96} by combining the local $\Ha$
objective prism survey results \cite{Gallegoetal95} and low
redshift ($z<1$) Canada-France redshift survey (CFRS)
results \cite{Lillyetal96} 
with estimates of the SFD at
$z\sim2.75$ and $z\sim4$ based on the Hubble Deep Field (HDF) 
data. 
These combined data sets indicate a SFD rapidly increasing with 
cosmic time at high redshift, reaching a maximum at $z\sim1.5-2$ and 
decreasing at lower redshifts \cite{Madauetal96}. 
Semi-analytical models of galaxy evolution in a
hierarchically clustering cold dark matter (CDM)
universe \cite{Baughetal97} provide a remarkably good fit to the
derived redshift dependence of the SFD.

This scenario provides a very suitable
starting point for future research. There are two obvious
lines of investigation which urgently need to be pursued and which can be
addressed by near-IR emission line surveys.

\begin{enumerate}
\item
Given the pronounced peak in the SFD between $z\sim1.5$ and 2,
and its remarkable agreement with the peak of the comoving
number density of active galactic nuclei (AGNs) at similar redshifts
\cite{Shaveretal96} and the peak in the density of damped $\Lya$ 
absorbers \cite{StorrieLombardietal96},
this epoch appears to be of fundamental 
importance for the formation of present-day
galaxies.
However, no direct measurements of the SFD in this redshift range are 
available, and the inferred peak in the SFD follows only from the fact
that both from higher and from lower
redshifts, the SFD increases as this epoch is approached. Direct
measurements
of the SFD out to $z\sim2.5$, i.e., covering the SFD peak era as well as
the preceding and following epochs, are therefore crucially important.
\item
The SFDs at $z\sim2.75$ and $z\sim4$ 
are based on the rest frame ultraviolet (UV) emission of
UV-bright
galaxies \cite{Madauetal96}. Therefore, 
these points are lower limits, since obscuration by dust will
give rise to an underestimate of the derived star formation rates
(SFRs) and since UV-obscured star forming galaxies will not be
accounted for in
the derived SFD\null. 
The colours of the
HDF objects used to derive the SFD at
$z\sim2.75$ and $z\sim4$ are similar to those
of local UV-selected, but dusty starburst galaxies and inconsistent with an
unreddened stellar population. 
Extinction corrections on the derived high-$z$ SFRs of more
than a factor of 10 are derived \cite{Meureretal97}. 
Obviously, the uncertainties inroduced by the presence of extinction
will be strongly alleviated if a star formation tracer at longer
wavelength can be used.
\end{enumerate}

\renewcommand{\arraystretch}{1.2}

\begin{table}[t]
{\bf Table 1.} Redshift ranges where various bright spectral lines are
available in near-IR atmospheric windows.\label{tab.Haz}
\begin{center}
\begin{tabular}{|c|c|c|c|c|c|}
\hline
Line & $\lambda$ & $Z$             & $J$             & $H$             & $K$ \\
 & [$\mu$m] & $0.85-0.99\mum$ & $1.00-1.34\mum$ & $1.50-1.80\mum$ & $2.03-2.35\mum$ \\
\hline
\hline
\multicolumn{6}{|c|}{\sl Good star formation tracers}\\
\hline
$\Ha$            & 0.6563 & $0.30-0.51$ & $0.52-1.04$ & $1.29-1.74$ & $2.09-2.58$ \\
$\rec{H}{b}$     & 0.4861 & $0.75-1.04$ & $1.06-1.76$ & $2.09-2.70$ & $3.18-3.83$ \\
$[\ion{O}{ii}]$  & 0.3727 & $1.28-1.66$ & $1.68-2.60$ & $3.02-3.83$ & $4.45-5.31$ \\
$\rec{Pa}{a}$    & 1.8751 & & & & $0.08-0.25$ \\
\hline
\hline
\multicolumn{6}{|c|}{\sl Poor star formation tracers}\\
\hline
$\Lya$           & 0.1215 & $6.00-7.15$ & $7.23-10.0$ & $11.4-13.8$ & $15.7-18.3$ \\
$[\ion{O}{iii}]$ & 0.5007 & $0.70-0.98$ & $1.00-1.68$ & $2.00-2.60$ & $3.05-3.69$ \\
\hline
\end{tabular}
\end{center}
\end{table}

Since $\Ha$ emission is much less strongly affected by dust than
shorter wavelength tracers (both in continuum and in spectral lines),
and since large format, high quality near-infrared arrays now
enable emission-line surveys in the $Z$, $J$, $H$ and $K$-bands,
$\Ha$ surveys will be able to address both of these points, and thus
supply unique information on the star
formation history of the universe out to $z\approx2.5$.

\mathversion{bold}

\section{Use and limitations of H$\alpha$ and other tracers}\label{sec.lines}

\mathversion{normal}

\subsection{Ly$\alpha$ is not a good tracer of star formation}

Traditionally,
surveys for high-$z$ star forming galaxies have targeted the $\rec{Ly}{a}$ 
line, which moves into the optical regime for $z=1.9-6.0$. However, since
$\rec{Ly}{a}$ is resonantly scattered, even very small quantities of dust will
effectively suppress the line 
\cite{CharlotFall91,CharlotFall93,ChenNeufeld94}. Indeed,
despite painstaking efforts, $\rec{Ly}{a}$ surveys have not revealed a 
population of high-$z$ starburst 
galaxies \cite{ThompsonDjorgovski95}. 
Recent spectroscopic observations of
a number of star forming $z>3$ galaxies selected by the UV dropout technique
\cite{Steideletal96a} confirm that $\rec{Ly}{a}$ is not a good tracer of star 
formation in high-$z$ galaxies. These galaxies form stars at rates of
$\sim10\Msun\pun{yr}{-1}$ as derived from their rest frame UV properties,
but $\rec{Ly}{a}$ is absent (or in {\it absorption}) in more than 50\% of 
the cases, while in most of the remaining objects the line is faint 
\cite{Steideletal96a}.
Thus, while $\rec{Ly}{a}$ searches can be used to find high-$z$ star forming
galaxies \cite{Macchettoetal93,Giavaliscoetal94,Huetal96}, 
such surveys will miss a large fraction of these
galaxies, and cannot provide reliable SFRs for the galaxies that are
detected.

\begin{table}[th]
{\bf Table 2.} 
Conversion factors of $\Ha$ luminosity $L_{{\rm H}\alpha}$ into SFR
determined for various IMFs and empirically, and into MER, 
and conversion factors of
various other line luminosities into $L_{{\rm H}\alpha}$, in
the sense Derived = Conversion $\cdot$ Observed.\label{tab.HaSFR}
\begin{center}
\begin{tabular}{|c|c|c|l|}
\hline
Derived & Conversion & Observed & Comments \\
\hline
\hline
\multicolumn{4}{|c|}{\sl Star formation rate $(\Msun\pun{yr}{-1})$
from $L_{{\rm H}\alpha}$ $(\Lsun)$}\\
\hline
SFR & $1.27\cdot10^{-8}$ & $L_{{\rm H}\alpha}$ & Salpeter IMF \cite{Salpeter55} \\
SFR & $4.06\cdot10^{-8}$ & $L_{{\rm H}\alpha}$ & Scalo IMF
\cite{Scalo86} \\
SFR & $1.83\cdot10^{-8}$ & $L_{{\rm H}\alpha}$ & Miller-Scalo IMF \cite{MillerScalo79} \\
SFR & $3.41\cdot10^{-8}$ & $L_{{\rm H}\alpha}$ & empirical
conversion \cite{Kennicutt83}\\
\hline
\hline
\multicolumn{4}{|c|}{\sl Metal ejection rate $(\Msun\pun{yr}{-1})$
from $L_{{\rm H}\alpha}$ $(\Lsun)$}\\
\hline
MER & $3.25\cdot10^{-10}$ & $L_{{\rm H}\alpha}$ & not sensitive to IMF \\
\hline
\hline
\multicolumn{4}{|c|}{\sl $L_{{\rm H}\alpha}$ $(\Lsun)$ from luminosity in other
lines $(\Lsun)$}\\
\hline
$L_{{\rm H}\alpha}$ & 2.75 & $L_{{\rm H}\beta}$ & sensitive to
extinction \\
$L_{{\rm H}\alpha}$ & 9.71 & $L_{{\rm Pa}\alpha}$ & \\
$L_{{\rm H}\alpha}$ & 2.22 & $L_{[{\rm O}\,\mbox{\sc ii}]}$ & 
empirical conversion \cite{Kennicutt92}, sensitive to abundance,\\[-1mm]
 & & & excitation, IMF and extinction \\
\hline
\end{tabular}
\end{center}
\end{table}
\nocite{Salpeter55}
\nocite{Scalo86}
\nocite{MillerScalo79}
\nocite{Kennicutt83}
\nocite{Kennicutt92}

\subsection{$\Ha$ as a tracer of star formation in galaxies}\label{sec.Ha}

The problems affecting $\Lya$ surveys 
are avoided by searching for $\rec{H}{a}$ emission
instead. The $\rec{H}{a}$ line is not resonantly scattered
and thus much less sensitive to the effects of small amounts of dust. In
addition, the broad-band 
extinction at $\rec{H}{a}$ ($6563\un{\mbox{\AA}}$) is much less than that
at $\rec{Ly}{a}$ ($1215\un{\mbox{\AA}}$), 
$A_{{\rm Ly}\alpha}/A_{{\rm H}\alpha}=4.28$ \cite{Cardellietal89}. 
Hence an $\rec{H}{a}$ survey should produce a
much more reliable measurement of the SFD of the high-$z$
universe.

In \tabref{tab.Haz} the redshift ranges where $\Ha$ and other lines
are available in
the near-IR windows are presented. Since the new $1024^2$ Rockwell
HgCdTe ``Hawaii'' arrays still have at least 50\% quantum efficiency at
wavelengths as short at $0.8\mum$, the $Z$-band ($0.85-0.99\mum$) is
included in this table.
For $\Ha$, the thermal background emission limits sensitive searches
to redshifts $z<2.5$.
Near-IR $\Ha$ surveys thus will be able to fill in the
gap between the $z<1$ results of the CFRS and $z>2.5$ results of the
HDF, and cover the crucial epoch where the SFD in the universe reaches its
maximum value. 

In interpreting $\Ha$ observations it is important to take into
account that, after correcting for extinction at $6563\un{\mbox{\AA}}$
in the rest frame, and under the usual assumption of case~B recombination in
ionization bounded, dust-free $\HII$ regions, $\Ha$ measures only the
production rate of hydrogen ionizing photons ($\lambda<912\un{\mbox{\AA}}$). 
Therefore, like the rest frame UV emission,
the $\Ha$ emission measures only the formation rate of {\it massive\/}
stars, and not directly the total SFR\null.
Hence, $\Ha$-derived SFRs are subject to the same uncertainties as UV-derived
SFRs as far as the conversion from {\it massive\/} SFR to {\it total}
SFR is concerned. This correction involves the {\it
assumption\/} of an Initial Mass Function (IMF) and is both large and
uncertain, as illustrated by \tabref{tab.HaSFR}.
However, like the UV emission, $\Ha$ emission can be used to reliably
measure a {\it metal
ejection rate\/} (MER), 
which is dominated by the same massive stars that power the
$\Ha$ emission. 

A final uncertainty is introduced by the presence of dust inside the
ionized regions. In evolved $\HII$ regions such as are found in the
disks of quiescent star forming galaxies like the Milky Way,
absorption by dust of ionizing photons is unimportant. In
(ultra)compact $\HII$ regions on the other hand, dust may absorb {\it
most\/} of
the ionizing photons. In Galactic ultracompact $\HII$
regions, at least 50\% 
of the ionizing radiation is absorbed by dust, and in
many cases more than 90\% \cite{WoodChurchwell89}. In these cases 
hydrogen recombination line
measurements, even at long wavelengths, will strongly underestimate
the Lyman continuum flux of the ionizing sources.
In luminous
starburst galaxies, and in particular in the more extreme cases such
as the ultraluminous infrared galaxies (ULIRGs), star formation takes
place in high density environments and most of the $\HII$ regions are
compact \cite{Zhaoetal97}. 
As a result, in the objects with the highest SFRs, $\Ha$
strongly underestimates the SFR\null. This problem cannot be solved by
observing longer wavelength recombination lines, since the line
emission is not absorbed but {\it quenched}, so that {\it all\/} 
nebular emission lines (and the radio free-free continuum)
will be suppressed.
This phenomenon is found in the prototypical ULIRG $\sou{Arp}{220}$ 
\cite{VanDerWerfIsrael97}. The low $\Brg$
recombination line flux of the nuclei of $\sou{Arp}{220}$ has been
attributed to a very high foreground extinction ($A_V\sim 50\mg$)
\cite{Sturmetal96}. However, this reasoning fails on the simple
grounds that nuclei of $\sou{Arp}{220}$ are directly detectable in the
$J$, $H$ and $K$-bands \cite{VanDerWerfIsrael97}, so that the
extinction towards the nuclei at these wavelengths must be relatively
small. Furthermore, the low radio free-free emission flux density of
the nuclei of $\sou{Arp}{220}$ \cite{Scovilleetal91} 
is unaccounted for in this model.
Absorption of ionizing photons by dust in the nuclei of
$\sou{Arp}{220}$ on the other hand naturally accounts for all of these
observations \cite{VanDerWerfIsrael97}.

This result has the important implication that shallow $\Ha$ surveys, probing 
only the most luminous objects are not suitable for measuring or even
constraining the SFD at any redshift. Reliable MERs and SFDs
can only be obtained if the luminosity function is probed down to the
level where typical local Sc galaxies become detectable, i.e., at
rest frame equivalent widths in the $\Ha+[\ion{N}{ii}]$
complex ${\rm EW}_0(\Ha+[\ion{N}{ii}])\ge30\un{\mbox{\AA}}$. At this
rest frame equivalent width level,
the survey depth becomes comparable to the local
$\Ha$ objective prism surveys \cite{Gallegoetal97}.  

\subsection{Use and misuse of other lines as star formation tracers}\label{sec.otherlines}

With the exception of $\Lya$, other hydrogen recombination lines than
$\Ha$ can in principle also be used as star formation tracers. Since
the ratios of these lines depend only weakly on physical consitions in
the photoionized regions, only extinction plays a role in determining
their relative strengths. Unextincted line ratios with respect to
$\Ha$ for the most relevant lines are given in
\tabref{tab.HaSFR}. The only line which can 
potentially produce {\it low}-redshift interlopers is $\rec{Pa}{a}$; at a
fixed observing wavelength, all
of the other lines sample higher redshifts than $\Ha$ and thus require
much more luminous (and hence rarer) objects to yield
detections. Since $\rec{Pa}{a}$ interlopers always occur at low $z$,
they are relatively easy to recognize. In contrast, optical surveys
are strongly plagued by low-$z$ interlopers in [$\ion{O}{ii}$] and
[$\ion{O}{iii}$] emission \cite{ThompsonDjorgovski95}.
Nevertheless, ine near-IR emission lines surveys it also remains necessary to 
spectroscopically identify the emission lines in all of
the objects detected, since these 
surveys are sensitive to {\it any\/} emission line redshifted to the
observing wavelength.
This property has often been used to artificially increase the
survey volume by simultaneously considering a number of
lines, assuming fixed ratios of these lines to $\Ha$
\cite{MannucciBeckwith95}. 
However, this procedure is both confusing and dangerous,
since ratios of e.g., [$\ion{O}{iii}$] $5007\un{\mbox{\AA}}$ 
to $\Ha$ in local galaxies
cover large ranges of values \cite{Kennicutt92}, 
and depend strongly on excitation and abundance effects.
Other than hydrogen recombination lines, only 
the [$\ion{O}{ii}$] $3727\un{\mbox{\AA}}$ line is a useable tracer of
star formation \cite{Kennicutt92}, although it is considerably less 
well-behaved than the hydrogen lines, and affected by abundance and
excitation effects, as well as extinction.
More fundamentally, results based on an emission-line survey at a 
fixed wavelength,
but calculated by considering {\it simultaneously\/} 
a number of different lines that
can be detected at this wavelength, thus
artificially increasing the instantaneous survey volume, completely
destroy any information on the redshift dependence of the SFD\null. The
result is a complicated weighted average value over a number of
redshifts, and very difficult to interpret.
Thus while an emission-line survey at a fixed wavelength can be used
to sample simultaneously volumes at a number of different redshifts in
different lines
(see \tabref{tab.Haz}), these volumes may not be combined to yield a 
``redshift-averaged'' SFD, but should be treated separately, providing
measurements at a number of different redshifts.

\section{Narrow-band imaging surveys}\label{sec.surveys}

Blind emission-line surveys are of three types, probing different
regions of parameters space \cite{Djorgovski92}: slitless
spectroscopy, long-slit spectroscopy, and
narrow-band imaging surveys. Because
of the bright background in the near-IR windows, which at
$\lambda<2.2\mum$ is dominated by many narrow airglow emission lines,
mainly due to OH,
near-IR surveys are best conducted using the narrow-band imaging method. 
Briefly, this technique consists of deep integration in a narrow-band
filter (optimally chosen such that bright airglow lines are avoided)
isolating a narrow redshift interval for a given spectral
line. Objects showing a narrow-band flux excess when compared to a
broad-band image are emission line candidates. Spectroscopic follow-up
for securely confirming and identifying the spectral lines and
for accurately determining their fluxes, is required. The choice of
narrow-band filter width is determined by a trade-off between survey
speed (which suffers if the filters are too narrow) and equivalent
width sensitivity (which is degraded for filters which are too wide).

\begin{figure}
\begin{center}
\epsfig{file=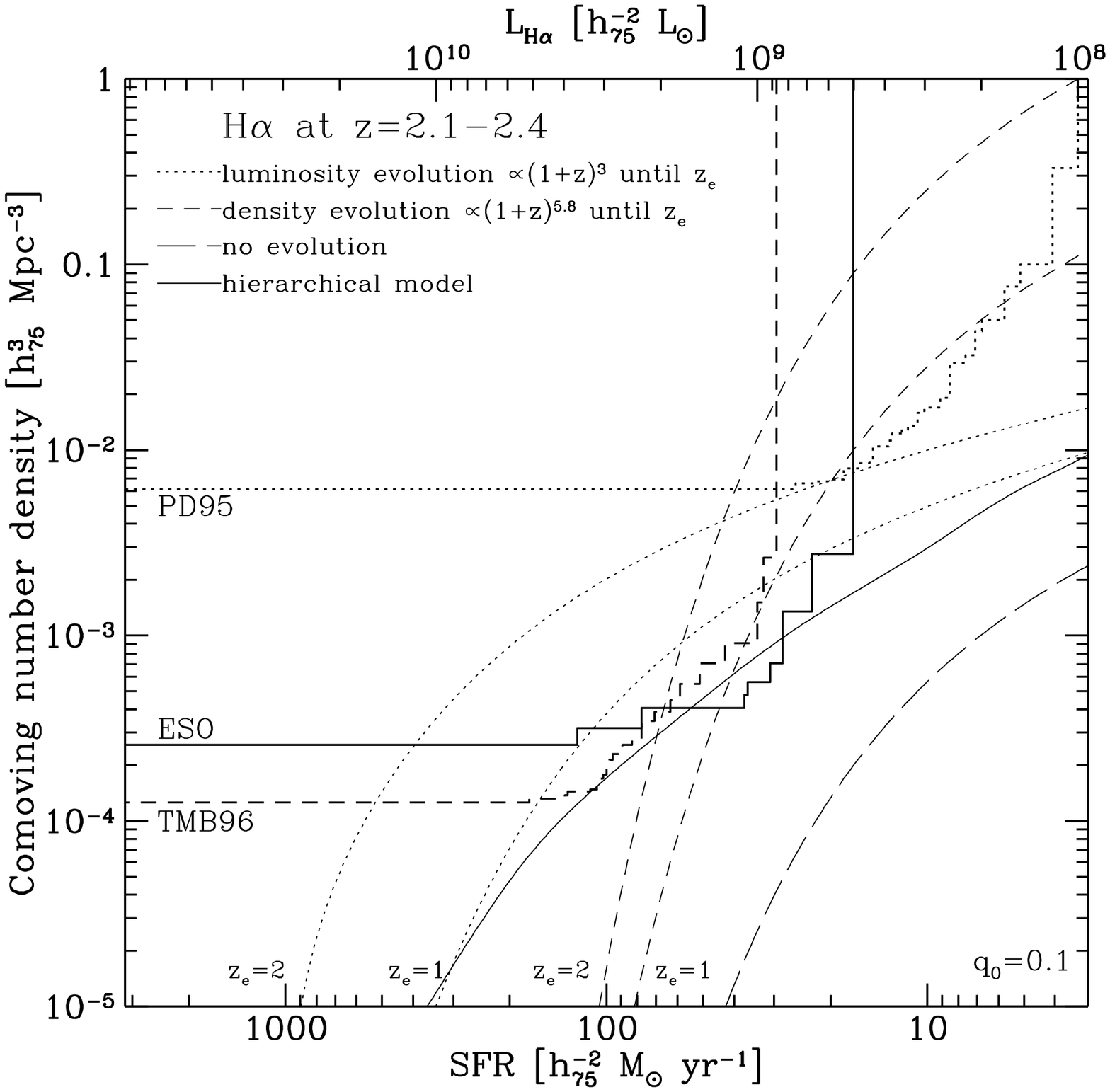,height=16cm}
\end{center}
{\bf Figure 1.}
Limits of the ESO near-IR $\Ha$ survey expressed as limiting 
comoving
number density (vertical axis) of galaxies with $\Ha$ luminosity 
(top horizontal axis) or total SFR (bottom horizontal axis) exceeding
the values indicated on the horizontal axes.
SFRs were calculated assuming the empirical conversion factor
\protect\cite{Kennicutt83} (see \protect\tabref{tab.HaSFR}).
The comoving number density limits represent 90\% confidence levels based on
Poisson statistics. $\Ha$ luminosity and SFR
limits denote $3\sigma$ levels.
For comparison, the results of two most important earlier surveys
\protect\cite{PahreDjorgovski95,Thompsonetal96}, 
reanalyzed in
a fashion identical to the ESO survey, and considering {\it
only\/} $\Ha$ at $z=2.1-2.4$ are also shown (labeled ``PD95'' 
and ``TMB96'' respectively). The long-dashed curve represents the 
cumulative local $\Ha$
luminosity function \protect\cite{Gallegoetal95}. 
The dotted curves show
this cumulative luminosity function after applying pure luminosity
evolution proportional
to $(1+z)^3$ ending at a redshift $\qu{z}{e}$. Similarly, the short-dashed
curves
are based on the local $\Ha$ luminosity function assuming pure density 
evolution proportional to $(1+z)^{5.8}$ ending at redshift $\qu{z}{e}$. 
Evolutionary calculations for both $\qu{z}{e}=1$ and $\qu{z}{e}=2$ are shown.
The continuous curve is the cumulative 
$\Ha$ luminosity function at $z=2.25$ resulting
from a semi-analytical calculation of galaxy
formation and evolution in a hierarchically clustering CDM universe 
\protect\cite{Baughetal97}.
\label{fig.lim}
\end{figure}

\subsection{The ESO near-IR $\Ha$ survey}\label{sec.lim}

Recently, a new survey 
has been carried out at the ESO/MPI $2.2\un{m}$ telescope 
\cite{VanDerWerfetal97}. Based on the considerations of
\secref{sec.Ha} the survey was designed to emphasize depth more than
area coverage. For the
first time, detections were obtained. 
A detailed description of the survey parameters and observation and
reduction procedures is given elsewhere \cite{VanDerWerfetal97}.
The total area covered is
$48.0\,\Box'$, and the total comoving volume
sampled for $\rec{H}{a}$ emission is
$9.0\cdot10^3\,h_{75}^3\pun{Mpc}{3}$ for $q_0=0.1$ (or
$3.0\cdot10^3\,h_{75}^3\pun{Mpc}{3}$ for $q_0=0.5$), which is about 50\% of
the volume of the largest survey to date 
\cite{Thompsonetal96}. However, with an
area-weighted $3\sigma$ point-source sensitivity of 
$1.9\cdot10^{-16}\esc\pun{\mum}{-1}$ 
in the narrow-band frame, the ESO survey goes considerably deeper. In the
entire survey, {\it two\/} serendipitous objects were found with a narrow-band
excess significant at about the $4\sigma$ level.

The limits implied by this survey are presented in \figref{fig.lim}, in terms
of a comoving number density of galaxies more luminous in $\Ha$ than
a certain luminosity.
The ESO survey produces
the best currently available limits for SFRs in the range of $20$ to 
$60\Msun\pun{yr}{-1}$. It is significant that the two serendipitously
detected objects found in
this survey have SFRs in this range. 
Locally, such SFRs are found in starburst 
galaxies.
It is clear that in the absence of evolution in the $\Ha$ luminosity 
function no detections would be expected in any of the surveys.
However, strong 
evolution is known to take place. Since both $\Ha$ and far-infrared
(far-IR) emission are
proportional to SFR, it is reasonable to assume similar evolution laws for
the $\Ha$ and far-IR luminosity functions. The evolution of the far-IR 
luminosity function can be described  by pure
luminosity evolution $\propto (1+z)^3$, or pure density evolution 
$\propto (1+z)^{5.8}$ \cite{RowanRobinson96}. 
The local $\Ha$ luminosity function, evolved according
to these evolutionary scenarios, is plotted in \figref{fig.lim}, where
we have continued the evolution to a redshift $\qu{z}{e}$. Inspection of
\figref{fig.lim} suggests that many detections would be expected for
$\qu{z}{e}=2$, and a fair number for $\qu{z}{e}=1$ as well. 
Since the highest SFRs will be underestimated by $\Ha$
(\secref{sec.Ha}), the analysis will be limited
to low luminosities.
For $\qu{z}{e}=2$, both luminosity and density evolution of the forms adopted
here are ruled out by the ESO survey; in the case of density evolution, the 
strongest constraint is provided by the small-area Keck survey 
\cite{PahreDjorgovski95}. For $\qu{z}{e}=1$, density evolution is
fully compatible with all surveys, while luminosity evolution
cannot be ruled out given that the $\Ha$
emission may be somewhat suppressed at the relevant SFRs. 
In summary, these results rule
out evolutionary scenarios of the types considered here
if this evolution continuous backward to $\qu{z}{e}=2$; however, for 
$\qu{z}{e}=1$ the survey limits are not violated. Obviously, the same kind
of agreement can be obtained by lowering the exponent of the evolutionary law.

A cumulative $\Ha$
luminosity function for $z=2.25$, kindly supplied by Carlton Baugh and 
based on a semi-analytical model of galaxy formation and
evolution in a hierarchically clustering CDM universe \cite{Baughetal97}, 
is also shown in \figref{fig.lim}. It is remarkable
that this model provides a very good match to the observational results.
This model predicts only detections in the ESO
survey, with SFRs of $30-70\Msun\pun{yr}{-1}$, in complete agreement
with the results. While the number of detections is still very small, 
it is clear that the ESO survey begins to approach
luminosities and volume densities predicted by sophisticated modeling. 

\begin{figure}
\begin{center}
\centerline{\epsfig{file=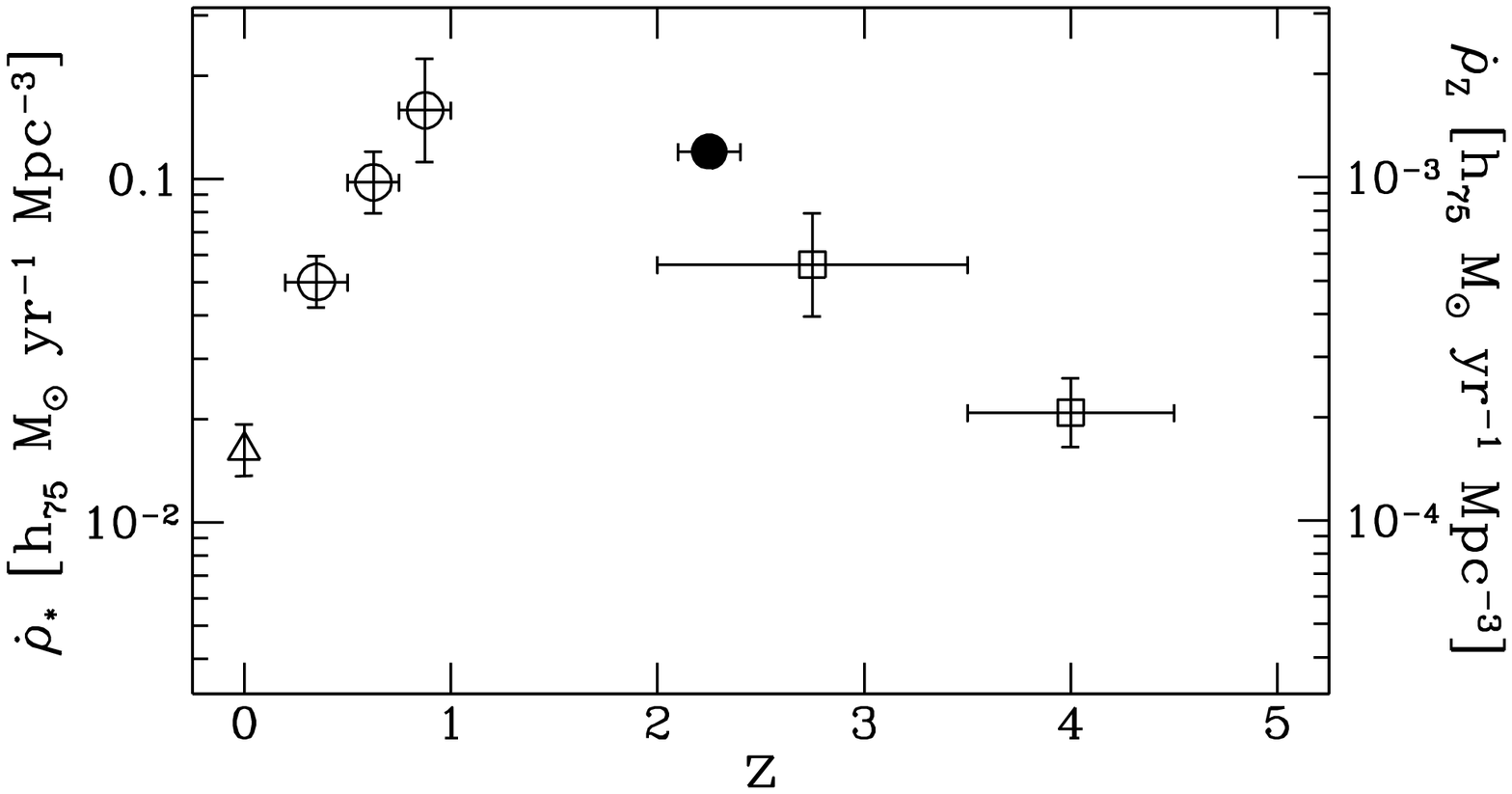,height=8cm,clip=}}
\end{center}
{\bf Figure 2.}
Implications of the $z=2.25$ results of the ESO near-IR
$\Ha$ survey (filled circle) for the
SFD (left vertical axis) and MER
(right vertical axis) compared to the rest frame UV-derived
values based on the CFRS \cite{Lillyetal96} (open circles) and the HDF
\cite{Madauetal96} (open squares) and the local $\Ha$ survey
\cite{Gallegoetal95} (open triangle).  The vertical error bar for the
ESO $\Ha$ point has been omitted.
Star formation densities were derived from $\Ha$
luminosity densities using the empirical conversion factor
\cite{Kennicutt83}. Points denoted by open symbols were converted to
the same scale using the conversion factors in \tabref{tab.HaSFR}. For
reasons discussed in \secref{sec.SFD}, the UV-derived points (open
circles and squares) should be considered lower limits.
\label{fig.SFD}
\end{figure}

\subsection{Implications for star formation at $z=2.25$}

Inspection of \figref{fig.lim} shows that only
part of the $\Ha$ luminosity function at $z=2.25$ is sampled. 
Hence, a SFD obtained by considering
only the detected objects is a firm 
{\it lower limit\/} to the total SFD at $z=2.25$. Not unexpectedly, 
the resulting value of 
$0.014\Msun\pun{yr}{-1}\pun{Mpc}{-3}$
lies significantly below the value expected based on rest frame UV data
\cite{Madauetal96}.

In order to estimate the {\it total\/} 
SFD at $z=2.25$, the part of the luminosity function
not represented in the survey must be accounted for. 
The detections are quite luminous and, in terms
of a Schechter luminosity function, likely significantly brighter
than $L_*$ at $z=2.25$. Consequently, the correction to a total SFD is large,
and, since $L_*$ at $z=2.25$ is not known, uncertain. As noted in 
\secref{sec.lim}, the luminosity function for $z=2.25$ based on the
semi-analytical hierarchical model \cite{Baughetal97} provides the best
fit to the available $\Ha$ data, and also correctly reproduces a number of
other observations. In the absence of a direct observational determination,
this luminosity function is therefore the best choice for $\Ha$ emission at 
$z=2.25$. Integration of this luminosity function yields a total comoving 
SFD at 
$z=2.25$ of $0.12\Msun\pun{yr}{-1}\pun{Mpc}{-3}$. As can be seen in 
\figref{fig.SFD}, this value is in good agreement with the SFD expected based
on the analysis by M96. Given the very small number of objects involved, the
agreement is very encouraging.

\section{Outlook}\label{sec.con}

The above results and considerations lead to two observational
recommendations.
\begin{enumerate}
\item In order to provide a measurement
of the SFD at a particular redshift, near-IR emission line surveys
must probe the luminosity function down to sufficiently faint
levels. Ideally, such surveys should aim to determine the luminosity
function of star forming galaxies at the relevant redshifts
directly, removing the need of assuming a luminosity function. 
The new generation of $8\un{m}$ class telescopes equipped
with large-format near-IR array cameras will provide the required
sensitivity and survey power. By carrying out narrow-band searches at
a number of discrete redshift intervals, selected to avoid bright
airglow lines in the narrow-band filters, near-IR $\Ha$ surveys will trace
the cosmic star formation history out to $z\sim2.5$, including the
important $z=1.5-2$ era where the cosmic star formation density is
believed to peak. Higher redshifts will be available by using
$\rec{H}{b}$ or [$\ion{O}{ii}$]. These surveys will be much less
sensitive to uncertainties and biases introduced by extinction than
rest frame UV based searches.
\item The brightest part of the galaxy luminosity function cannot be
probed using nebular emission line surveys. Here
total SFRs must be derived
from the rest frame far-IR continuum. At low $z$, the brightest, optically 
obscured part of the
luminosity function accounts for only a small fraction of the total
SFD\null. However, if strong luminosity evolution takes place, this
fraction may become significant at higher redshift.
Blind submillimetre surveys (sampling the far-IR continuum of
high-$z$ galaxies) are required to address this point
observationally. The FIRST mission will play a major role in this
field. In the immediate future, a first assessment of the importance
of dusty, very luminous star forming galaxies at high $z$ will be made
with the SCUBA submillimetre camera at the James Clerk Maxwell Telescope.
\end{enumerate}

\acknowledgements{
I am very grateful to Nicolas Cretton for translating the abtract of
this review into French, and to Carlton Baugh for
for supplying the cumulative luminosity function shown
in \figref{fig.lim} based on his hierarchical model. Of those involved in the 
ESO near-IR $\Ha$ survey, I would like to thank in particular Alan
Moorwood and Malcolm Bremer, for their efforts and for ongoing
discussion. Finally, I would like to thank the organizers 
for inviting me to this very enjoyable meeting.
The research of Van der Werf has been made possible by a fellowship of
the Royal Netherlands Academy of Arts and Sciences.
}

\vfill
\end{document}